\documentclass[hyper]{JHEP} 

\usepackage{epsfig}
\usepackage{graphicx}

\def\be{\begin{equation}}
\def\ee{\end{equation}}
\def\ba{\begin{array}}
\def\ea{\end{array}}
\def\bea{\begin{eqnarray}}
\def\eea{\end{eqnarray}}
\def\nn{\nonumber\\}

\def\ba{\mathbf{a}}

\def\ph{\phi}

\def\ps{\psi}

\def\l{\lambda}

\def\th{\theta}

\def\r{\rho}

\def\tR{\tilde{R}}

\def\half{\frac{1}{2}}

\def\ls{\left(}

\def\rs{\right)}

\def\det{{\rm det}}

\newcommand\fverb{\setbox\pippobox=\hbox\bgroup\verb}

\newcommand\fverbdo{\egroup\medskip\noindent%

            \fbox{\unhbox\pippobox}\ }

\newcommand\fverbit{\egroup\item[\fbox{\unhbox\pippobox}]}

\newbox\pippobox

\title{Wilson Loops  in $3d~
QFT$ from D-branes in AdS$_4 \times
{\bf CP}^3$}
\author{J. Kluso\v{n}\\
Department of
Theoretical Physics and Astrophysics\\
Faculty of Science, Masaryk University\\
Kotl\'{a}\v{r}sk\'{a} 2, 611 37, Brno\\
Czech Republic\\

E-mail: \email{klu@physics.muni.cz}}

\author{Kamal L. Panigrahi\\
Department of Physics\\
Indian Institute of Technology Guwahati, \\
Guwahati , 781 039, INDIA \\
E-mail: \email{panigrahi@iitg.ernet.in}}

\preprint{\hepth{}}

\abstract{We study the Wilson loops  in
the three dimensional QFT from the
D-branes in the AdS$_4 \times$ CP$^3$
geometry. We find out explicit D-brane
configurations in the bulk which
correspond to both straight and
circular Wilson lines extended to the
boundary of AdS$_4$. We analyze
critically the role of boundary
contributions to the D2-branes with
various topology and fundamental string
actions. }

\def \ba{\mathbf{a}}
\def\tr{\mathrm{Tr}}

\def\bA{\mathbf{A}}

\def\mF{\mathcal{F}}

\def\mL{\mathcal{L}}

\def\tR{\tilde{R}}

\begin{document}

\section{Introduction and Summary}\label{first}
Recently, Aharony, Bergman, Jafferis and Maldacena (ABJM)
\cite{Aharony:2008ug} have proposed a new class of gauge-string
duality between ${\cal N} = 6$ Chern-Simons theory and type IIA
string theory on AdS$_4 \times CP^3$. More precisely,  ABJM theory
has been conjectured to be dual to M-theory on AdS$_4 \times
S^7/Z_k$ with $N$ units of four-form flux which for $1 << N <<
k^4$ can be compactified to type IIA theory on AdS$_4 \times
CP^3$, where $k$ is the level of Chern-Simon theory with gauge
group $SU(N)$. The ABJM theory is weakly coupled for $\lambda <<
1$, where $\lambda = N/k$ is the 't Hooft coupling.

In proving AdS$_5$/CFT$_4$ duality, the integrability of both the
string and the gauge theory side has played a key role. The
semiclassical string states in the bulk side has been used to look
for suitable gauge theory operators in the dual side, in
establishing the correspondence. This important observation makes
one to believe that perhaps a similar structure of integrability
can be employed in the recently proposed AdS$_4$ /CFT$_3$ to
understand it better. Indeed in \cite{Minahan:2008hf,Gaiotto:2008cg,
Gromov:2008qe}
there has been attempts
along this and it seems quite interesting to study the
semiclassical rotating strings in particular sectors of the
theory (see for example \cite{Grignani:2008te}-\cite{Gromov:2008fy}).
For example, the giant magnon solution and spike string
solutions has been studied and they would certainly correspond to
the trace operators in the three dimensional CFT\footnote{for more related
work see \cite{Benna:2008zy}-\cite{Rashkov:2008rm}}.

Further in the gauge theory, Wilson
loop operators are  non-local gauge
invariant operators in gauge theory in
which the theory can be formulated. In
the absence of matter, the Wilson loops
in Chern-Simons theory compute
topological objects as knot invariants
\cite{Witten:1988hf}, and are somewhat
less interesting than in four
dimensions, where they can be used as
an order parameter for confinement. For
theories coupled to matter, on the
other hand, we expect to find a similar
structure to the four-dimensional case
of N = 4 SYM, where the definition of
these operators involves the scalar
fields in a non-trivial way. In fact,
one  defines a Wilson loop as the trace
in an arbitrary representation $R$ of
the gauge group $G$ of the holonomy
matrix associated with parallel
transport along a closed curve $C$ in
spacetime. Since the beginning of the
proposed AdS/CFT correspondence
\cite{Maldacena:1997re}, it is known
that Wilson loops in $N=4$ SYM theory
can be calculated in dual description
using macroscopic strings
\cite{Maldacena:1998im,Rey:1998ik}.
This prescription is based on a picture
of the fundamental string ending on the
boundary of AdS$_5$ along the path $C$
specified by the Wilson loop operator.
The description of this Wilson loop in
terms of a fundamental string is a well
established part of the AdS/CFT
dictionary. In a recent interesting
paper \cite{Ishizeki:2008dw} a class of
new open string solutions in $AdS_5$
were found which end at the boundary on
various Wilson lines. It was found that
these configurations arising out of the
solutions to the equations of motion
corresponding to fundamental strings,
they describe Wilson loops in the
fundamental representations.

Motivated by the recent development of
AdS$_4$/CFT$_3$ duality, we would like
to find out these Wilson line solutions
in the AdS bulk, which will correspond
to non-local gauge invariant operators
in the CFT side. However, we will look
at various D-branes in the AdS side and
argue for the Wilson loop solutions.
Some time back it was argued in very
interesting paper \cite{Gomis:2006sb},
for type IIB theory, that Wilson loops
have a gravitational dual description
in terms of D5-branes or alternatively
in terms of D3-branes in $AdS_5\times
S^5$ background \footnote{For closely
related works, see
\cite{Bonelli:2008rv,Chen:2008ds,Drukker:2007qr,Lunin:2007zz,Sakaguchi:2007ea,
Chu:2007pb,Chen:2007ir,Gomis:2006mv,Drukker:2006zk,Gomis:2006im,
Armoni:2006ux,Hartnoll:2006ib,Drukker:2006ga,Lunin:2006xr,Hartnoll:2006hr}.}.
More precisely, in
\cite{Drukker:2005kx} it was argued
that a Wilson loop with matter in the
rank $l$ \emph{symmetric
representation} is better described as
a D3-brane embedded in $AdS_5$ with $l$
units of electric flux. Further, it was
argued in
\cite{Hartnoll:2006hr,Hartnoll:2006ib,Yamaguchi:2006tq}
that a Wilson loop with matter in the
rank $l$ \emph{antisymmetric
representation} is better described by
a D5-brane whose world-volume is a
minimal surface in the AdS part of the
geometry times an S$^4$ inside the
$S^5$ and that has a support from
$l$-units of world-volume electric
flux.

Hence, it seems that perhaps there are
D-brane representations in type IIA
theory which correspond to Wilson lines
in the dual 3$d$ CFT. In fact various
particle like branes has been found out
in \cite{Aharony:2008ug} wrapping
various cycles in CP$^3$. There are
D0-branes, D2-brane wrapped $CP^1
\subset CP^3$, D4-branes wrapped on
$CP^2 \subset CP^3$ and D6-brane
wrapped on $CP^3$ and so on. We will
however, find out solutions for the
D-branes which correspond to Wilson
lines in the $3d$ boundary theory. We
study the D2-brane completely in the
Euclidean AdS$_4$ in analogy with the
recent paper \cite{Drukker:2008wr}.
Moreover, by  adding a surface term, in
accordance with \cite{Drukker:2005kx}
we find that this solution has similar
form as D3-brane configuration in
$AdS_5$ and it   corresponds to line
operator in $N=4$ SYM theory. We also
argue that the induced geometry on the
world-volume of D2-brane is
$AdS_2\times S^1$. Moreover we also
show that the action evaluated on the
classical configuration vanishes and
hence we can expect that the vacuum
expectation value of dual line operator
in $3d$ theory is equal to one.
 Then
our results can be interpreted as a
support for an existence of line
operators in $3d$ theory. It would be
really interesting to find their
explicit form.

As the next step we propose another
D2-brane configurations with  topology
AdS$_2 \times$ S$^1$, corresponding to
the straight and circular Wilson lines.
These after adding the boundary terms
to cancel the divergence of the action
while evaluated on a straight surface
have shown to behave like Wilson lines
that end on the boundary of AdS.

The rest of the paper is organized as
follows. In next section (\ref{second})
we review basic facts about ABJM
theory. Then in section (\ref{third})
we review description of Wilson loops
using fundamental string in dual
geometry. Then in section
(\ref{fourth}) we study the D2-brane
inside the Euclidean AdS$_4$. We show
that this configuration corresponds to
the dual surface operator of dimension
one. Then we study the D2-brane
configuration which has AdS$_2 \times$
S$^1$ topology. We find the D2-branes
which correspond to straight and
circular Wilson line solutions. We
study the boundary terms in both the
cases. We conclude in section
(\ref{fifth}).

\section{Basic Facts About ABJM
Theory}\label{second}
 The purpose of
this section is to outline the basic
facts about the ABJM theory. It is a
$3d$ superconformal Chern-Simons-matter
theory with explicit $N=6$
supersymmetry and it can be interpreted
as a theory describing $N$ coincident
M2-branes at the singularity of the
orbifold $C^4/Z_k$. This theory has
following basic properties:
\begin{itemize}
\item Gauge and global symmetries:
\begin{eqnarray}
& &\mathrm{gauge \ symmetry:} \quad
U(N)\times \overline{U(N)} \ , \nonumber \\
& &\mathrm{global \ symmetry:} \quad
SU(4)
\ .  \nonumber \\
\end{eqnarray}
We also denote trace over $U(N)$ and
$\overline{U(N)}$ as $\tr$ and
$\overline{\tr}$ respectively.
\item
The on-shell fields are gauge fields,
together with complexified Hermitian
scalars and Majorana spinors
$(A=1,2,3,4)$:
\begin{eqnarray}
A_\mu: & & \quad \mathrm{Adj}(U(N)) \ ,
\quad \overline{A}_\mu: \quad
\mathrm{Adj}(\overline{U(N)}) \ , \quad
\mu,\nu=0,1,2 \ , \nonumber \\
Y^A
&=&(X^1+iX^5,X^2+iX^6,X^3-iX^7,X^4-iX^8):
\quad \mathbf{(N,\overline{N};4)}
\ , \nonumber \\
Y^\dag_A
&=&(X^1-iX^5,X^2-iX^6,X^3+iX^7,X^4+iX^8):
\quad
\mathbf{(\overline{N},N,\overline{4})}
\ , \nonumber \\
\Psi_A
&=&(\psi^2+i\chi^2,-\psi^1-i\chi^1,\psi_4-i\chi_4,
-\psi_3+i\chi_3): \quad
\mathbf{(N,\overline{N};4)} \ ,
\nonumber
\\
\Psi^{\dag A}&=&
(\psi_2-i\chi_2,-\psi_1+i\chi_1,\psi^4+i\chi^4,
-\psi^3-i\chi^3): \quad
\mathbf{\overline{N},N;4)} \ .  \nonumber \\
\end{eqnarray}
\item The action of ABJM theory takes
the form ($\kappa=\frac{k}{2\pi}$)
\begin{eqnarray}
S&=&\kappa \int d^3x
\left[\epsilon^{\mu\nu\lambda}
\tr\left(\frac{1}{2}A_\mu\partial_\nu
A_\lambda+\frac{i}{3}A_\mu A_\nu
A_\lambda\right)-\epsilon^{\mu\nu\lambda}
\overline{\tr}\left(\frac{1}{2}
\overline{A}_\mu\partial_\nu\overline{A}_\lambda+
\frac{i}{3}\overline{A}_\mu
\overline{A}_\nu\overline{A}_\lambda\right)+
\right.
\nonumber \\
&+&\frac{1}{2}\overline{\tr}
\left(-(D_\mu Y)_A^\dag D^\mu
Y^A+i\overline{\Psi}^{\dag A} \gamma
^\mu D_\mu \Psi_A\right)+ \frac{1}{2}
\tr\left(-D_\mu Y^A (D^\mu Y)^\dag_A
+i\overline{\Psi}_A \gamma^\mu D_\mu
\Psi^{\dag A}\right)- \nonumber
\\
&-&\left.V_F-V_B\right] \ ,   \nonumber \\
\end{eqnarray}
where the covariant derivatives are
defined as
\begin{equation}
D_\mu Y^A=
\partial_\mu Y^A+iA_\mu Y^A-
i Y^A \overline{A}_\mu \ , \quad (D_\mu
Y)^\dag_A=
\partial_\mu Y^\dag_A+i\overline{A}_\mu
Y^\dag_A-iY^\dag_A A_\mu
\end{equation}
and similarly for fermions
$\Psi_A,\Psi^{\dag A}$.  Finally, the
potential terms are
\begin{eqnarray}
V_F&=&i\overline{\tr} \left[ Y_A^\dag
Y^A \Psi^{\dag B}\Psi_B-2Y^\dag_AY^B
\Psi^{\dag A}\Psi_B+\epsilon^{ABCD}
Y^\dag_A \Psi_B Y^\dag_C\Psi_D\right]-
\nonumber \\
&-&i\tr\left[Y^A Y_A^\dag \Psi_B
\Psi^{\dag B}-2Y^A Y_B^\dag \Psi_A
\Psi^{\dag B}+\epsilon_{ABCD}
Y^A\Psi^{\dag B}Y^C \Psi^{\dag
D}\right] \nonumber \\
\end{eqnarray}
and
\begin{eqnarray}
V_B&=&-\frac{1}{3}\overline{\tr} \left[
Y^\dag_A Y^B Y_B^\dag Y^C Y^\dag_C Y^A
+Y^\dag_A Y^A Y^\dag_B Y^B Y^\dag_C
Y^C+\right. \nonumber \\
&+&\left. 4 Y_A^\dag Y^B Y_C^\dag Y^A
Y_B^\dag Y^C-6 Y^\dag_A Y^A Y^\dag_B
Y^C Y_C^\dag Y^A\right] \nonumber \\
\end{eqnarray}
\end{itemize}

\section{Fundamental string as Wilson
line}\label{third}

 In this section we review the
description of Wilson loop using
fundamental string in the bulk of
$AdS_4$. We are interested in the
configuration when string sigma model
is embedded in $AdS_4$ only. Since
$AdS_4$ can be trivially embedded in
$AdS_5$ it is clear that  many results
that were derived for Wilson loops in $
N = 4$ SYM are valid also for  Wilson
loops in $N = 6$ CS.

We begin with Nambu-Gotto action
\begin{equation}\label{NGaction}
S_F=\tau_F \int d\tau d\sigma
\sqrt{\det \ba_{\mu\nu}} \ ,
\end{equation}
where
\begin{equation}
\ba_{\mu\nu}=\partial_\mu X^m
\partial_\nu X^n g_{mn} \ , \quad
\mu,\nu=\tau,\sigma \ .
\end{equation}
and where $\tau_F=\frac{1}{2\pi}$ in
our units.
In order to describe infinite Wilson
line we consider an ansatz
\begin{equation}
X^0=\tau \ , \quad  Y=\sigma \ , \quad
X^1\equiv X=X(\sigma)
\end{equation}
while remaining fields are constant.
For this ansatz we have
\begin{equation}
\ba_{\tau\tau}=\frac{\tR^2}{\sigma^2} \
, \quad \ba_{\sigma\sigma}=g_{yy}
\partial_\sigma Y\partial_\sigma Y+
g_{xx}\partial_\sigma X\partial_\sigma
X= \frac{\tR^2}{\sigma^2} ( 1+X'^2)
\end{equation}
and hence the action (\ref{NGaction})
 takes the form
\begin{equation}
S_F=\tau_F\tR^2\int d\tau d\sigma
\frac{1}{\sigma^2}\sqrt{1+X'^2} \ .
\end{equation}
Further, the equation of motion for $X$
takes the form
\begin{equation}
\partial_\sigma
[\frac{X'}{\sigma^2\sqrt{1+X'^2}}]=0
\end{equation}
that has clearly solution as
$X'=0\Rightarrow X=const$. Then this
ansatz above describes infinite long
Wilson line. For this line the action
takes the form
\begin{eqnarray}
S_F=\tau_{F}\tR^2\int_{-T/2}^{T/2}
d\tau \int d\sigma
\frac{1}{\sigma^2}\sqrt{1+X'^2}= \tau_F
\tR^2 T \int_\epsilon^\infty d\sigma
\frac{1}{\sigma^2}= \tau_F T
\tR^2\frac{1}{\epsilon} \ ,  \nonumber \\
\end{eqnarray}
where in order to derive finite result
we  imposed condition that Wilson line
is extended in time interval
$(-T/2,T/2)$.

 Let us add boundary term
to the action. Recall that we have
firstly evaluate this term for
arbitrary $Y$ and $X$, then include the
ansatz given above and finally evaluate
it on the surface $Y=\sigma=\epsilon$
\begin{eqnarray}
\delta S_F&=&-\int_{-T/2}^{T/2} d\tau
\frac{\delta \mL}{\delta\partial_\sigma
Y} Y=
\nonumber \\
&=&-\tau_F T  g_{yy}\partial_\sigma Y
Y\frac{\sqrt{g_{00}(\partial_\tau
X^0)^2}}{\sqrt{ g_{yy}(\partial_\sigma
Y)^2+g_{xx}(\partial_\sigma X)^2}}
= -\tau_F T
\tR^2\frac{1}{\epsilon} \nonumber \\
\end{eqnarray}
and we obtain familiar result that
\begin{equation}
S_F+\delta S_F=0 \
\end{equation}
that shows that the expectation value
of corresponding Wilson line is $<W>=1$
so that this Wilson line preserves some
fraction of supersymmetry. We discuss
the number of unbroken supersymmetries
below.

It is important property of
superconformal theories that straight
Wilson loop can be conformally
transformed to space-like circular
Wilson loop that has following string
theory description. To begin with let
us consider  the metric in the form
\begin{equation}
ds^2=\frac{\tR^2}{y^2}(dy^2+dr^2+r^2
d\phi^2+d(x^0)^2) \ .
\end{equation}
Then we  consider an ansatz
\begin{equation}
Y=\sigma \ , \quad X^0=const, \quad
\tau=\phi \ , R=R(\sigma) \ .
\end{equation}
Then
\begin{equation}
\ba_{\sigma\sigma}=\frac{\tR^2}{y^2}
((\partial_\sigma Y)^2+(\partial_\sigma
R)^2) \ , \quad
\ba_{\tau\tau}=\frac{\tR^2R^2}{Y^2}
\end{equation}
and hence the action (\ref{NGaction})
takes the form
\begin{equation}
S_F=\tau_F\tR^2\int d\tau d\sigma
\frac{R}{Y^2} \sqrt{(\partial_\sigma
Y)^2+(\partial_\sigma R)^2} \ .
\end{equation}
>From this action we easily determine
the equation of motion for $R$
\begin{eqnarray}\label{eqc}
\frac{1}{Y^2}\sqrt{1+(\partial_\sigma
R)^2} -\partial_\sigma[\frac{R}{Y^2}
\frac{\partial_\sigma
R}{\sqrt{1+(\partial_\sigma R)^2}}]=0 \
.
\nonumber \\
\end{eqnarray}
Then it can be easily shown that
(\ref{eqc}) can be solved with the
ansatz
\begin{equation}\label{RKs}
R^2=K^2-\sigma^2 \ ,
\end{equation}
for constant $K$.
%
Let us now evaluate the action for the
solution (\ref{RKs})
\begin{eqnarray}
S_F=\tau_F \tR^2\int_{0}^{2\pi} d\tau
\int_\epsilon^K d\sigma
\frac{R}{\sigma^2}
\sqrt{1+(\partial_\sigma R)^2}= -2\pi
\tau_F \tR^2
K[\frac{1}{K}-\frac{1}{\epsilon}] \ .
\nonumber \\
\end{eqnarray}
Further, we add to the action the
boundary term in the form
\begin{eqnarray}
\delta S_F=-\int_0^{2\pi} d\tau
\frac{\delta \mL} {\delta_\sigma Y}Y= -
2\pi\tau_F \tR^2\frac{R}{Y^2}
\frac{\partial_\sigma YY}
{\sqrt{(\partial_\sigma
Y)^2+(\partial_\sigma R)^2}}=
-2\pi\tau_F \frac{K}{\epsilon}\nonumber
\\
\end{eqnarray}
Then we obtain that the whole action is
finite and equal to
\begin{equation}
S_F+\delta S_F= -2\pi \tau_F \tR^2=
-\pi\sqrt{2\lambda} \ .
\end{equation}
using the fact that
$\tau_F=\frac{1}{2\pi}$ and
$\tR^2=\pi\sqrt{2\lambda}$.

Let us now briefly review the
explanation why the expectation values
of the straight and circular Wilson
lines are different
\cite{Drukker:2000rr}, at least in case
of $CFT_4/AdS_5$ correspondence. As was
argued there the origin in the
difference is in the application of the
conformal transformation that maps line
to the circle. In fact, in order to map
the line to the circle we have to add
the point at infinity to the line. Then
there is a small difference in the
calculation of the perturbative theory
when we have to add a total derivative
to the propagator. Then it was shown in
\cite{Drukker:2000rr} that the
perturbative calculation on SYM side
agrees with the string theory
description of this Wilson line.

Let us now briefly discuss the
space-time symmetries that are
preserved by these classical string
solutions. It can be shown that each
string configuration wraps an
appropriate $AdS_2$ submanifold in
$AdS_4$ so that it preserves
$SL(2,R)\times SO(2)$ symmetry of
isometry $SO(2,3)$ of $AdS_4$. On the
other hand the string is localized at
$CP^3$ so that it breaks the original
isometry $U(4)$ of $CP^3$ to
$U(1)\times U(3)$.  Further, it can be
shown that both these string
configurations preserve $12$
supercharges from $24$ supercharges of
original background.

With analogy with $AdS_5/CFT_4$
correspondence we suggest that this
string configuration is dual straight
and circular Wilson line in dual
theory. However we leave the analysis
of the properties of these objects for
further works.

\section{D-branes in Euclidean AdS$_4 \times$ CP$^3$}
\label{fourth}
 We start by writing down
the metric for $AdS_4 \times CP^3$,
which in a particular parametrization
reads
\bea ds^2 &=& \tR^2( ds_{AdS_4}^2+4 ds_{CP_3}^2)\nonumber \\
  ds_{AdS_4}^2 &=& - \cosh^2 \r \
dt^2 + d\r^2 + \sinh^2 \r \ls d \th^2 +
\sin^2 \th d \ph^2 \rs   \nn
ds_{CP_4}^2&=&  d\xi^2 + \cos^2 \xi
\sin^2 \xi \ls d \ps + \half \cos \th_1
d \ph_1 - \half \cos \th_2 d \ph_2
\rs^2  \nn  \ \ & &  + \frac{1}{4}
\cos^2 \xi \ls d \th_1^2 + \sin^2 \th_1
d \ph_1^2 \rs + \frac{1}{4} \sin^2 \xi
\ls d \th_2^2 + \sin^2 \th_2 d \ph_2^2
\rs  , \eea where
\begin{equation}
0\leq \xi \leq \frac{\pi}{2} \ , \quad
0\leq \phi_i\leq 2\pi \ , \quad 0\leq
\theta_i \leq \pi \ , \quad , i=1,2 \ ,
\end{equation}
and where
\begin{equation}
\tR^2=\frac{R^3}{4k} \ , \quad
e^{2\Phi_0}=\frac{R^3}{k^3} \ .
\end{equation}
The 't Hooft coupling constant is $\l
\equiv N/k$ where $k$ is the level of
the 3-dimensional N=6 ABJM model. The
relation between the parameters of the
string background and of the field
theory are (for $\alpha' = 1$)
\begin{equation}
\tR^2=\pi\sqrt{\frac{2N}{k}}=
\pi\sqrt{2\lambda} \ .
\end{equation}
At the same time the two-form field
strength is given by
\begin{eqnarray}
F^{(2)} = k(-\cos\xi \sin \xi
d\xi\wedge (2d\psi+\cos\theta_1
d\phi_1-\cos\theta_2 d\phi_2)-
\nonumber \\
 - \frac{k}{2}\left(\cos^2 \xi
\sin \theta_1 d\theta_1 \wedge d\phi_1
+ \sin^2 \xi \sin \theta_2 d\theta_2
\wedge d\phi_2\right) \nonumber \\
 \end{eqnarray}
and the four-form field
\begin{equation}
F^{(4)}=\frac{3R^3}{8} d\Omega_{AdS_4} \ ,
\end{equation}
where $d\Omega_{AdS_4}$ is unit volume
form of the $AdS_4$. There exist
freedom in determination of the
three-form $C^{(3)}$ and we choose
following one
\begin{equation}\label{C3}
C^{(3)}=\frac{R^3}{8}\frac{1}{y^3}
dx^0\wedge dx^1\wedge dx^2 \ .
\end{equation}

The dynamics  of Dp-brane in general background is governed by
following Dirac-Born-Infeld type of action including the
Wess-Zumino term:
\begin{eqnarray}\label{actD1}
S&=&S_{DBI}+S_{WZ} \ , \nonumber \\
S_{DBI}&=&-\tau_p \int d^{p+1}\zeta e^{-\Phi}
\sqrt{-\det\bA} \ , \nonumber \\
\bA_{\alpha\beta}&=&\partial_\alpha x^M\partial_\beta x^N
G_{MN}+(2\pi\alpha')\mF_{\alpha\beta}  \ , \nonumber \\
  \mF_{\alpha\beta}&=&
\partial_\alpha A_\beta-\partial_\beta A_\alpha-
(2\pi\alpha')^{-1}B_{MN}\partial_\alpha x^M\partial_\beta x^N
 \ , \nonumber \\
S_{WZ}&=&\tau_p\int e^{(2\pi\alpha')\mF}\wedge C \ ,  \nonumber \\
\end{eqnarray}
where $\tau_p$ is D$p$-brane tension,
$\xi^\alpha,\alpha=0, 1,\dots ,p$ are
the $(p+1)$ world-volume coordinates
and where $A_\alpha$ is gauge field
living on the world-volume of
D$p$-brane. Note also that $C$ in the
last line in (\ref{actD1}) means
collection of Ramond-Ramond fields.

\subsection{D2-brane}
Our goal is to find D-brane description
of Wilson lines in dual 3d theory. In
order to do this we will consider
Euclidean version of $AdS_4$ and also
write  it in the following form
\begin{equation}
ds^2_{AdS_4}=\frac{\tR^2}{ y^2}[dy^2+(dx^\mu)^2]=
\frac{\tR^2}{y^2}[{(dx^0)}^2 + dy^2+dr^2+r^2d\alpha^2]
 \ , \quad
\mu=0,1,2 \ ,
\end{equation}
where the boundary of $AdS_4$ is at
$y=0$. Let us consider following
D2-brane configuration
\begin{eqnarray}\label{defect}
x^0&=&\xi^0 \ , \quad  r=\xi^1  \ ,
\quad \alpha=\xi^2 \ ,  \quad y=y(r) \
, \nonumber \\
\xi &=& \mathrm{const}, \quad
\theta_i=\mathrm{const}, \quad
\phi_2=\mathrm{const} \ , \quad
\phi_1=\phi_1(\alpha) \ , \nonumber \\
\end{eqnarray}
where $\xi, \theta_1, \theta_2, \phi_1, \phi_2$ are the
coordinates of CP$^3$. This configuration should correspond to a
topological two dimensional operator in the dual CFT on the
boundary of the AdS$_4$.

Further, for the ansatz (\ref{defect})
the matrix $\bA$ takes the form
\begin{eqnarray}
\bA_{00}&=&\frac{\tR^2}{y^2} \ , \quad
\bA_{rr}= \frac{\tR^2}{y^2}(1+y'^2) \ ,
\nonumber \\
\bA_{\alpha\alpha}&=&\frac{\tR^2}{y^2}\left(1+
y^2(\cos^2\xi\sin^2\xi\cos^2\theta_1
+\cos^2\xi\sin^2\theta_1)\dot{\phi}_1^2\right)
\ , \nonumber \\
\end{eqnarray}
where $y'=\frac{dy}{dr} \ ,
\dot{\phi}_1=\frac{d\phi_1}{d\alpha}$.
Then it is easy to see that the
D2-brane action takes the form
\begin{eqnarray}\label{SD21}
&&S_{D2}= \tau_{2}\int d^3\zeta e^{-\Phi_0}\sqrt{\det
\bA}-\tau_{2}\int C^{(3)}=\nonumber \\
&=& \tau_2  \frac{R^3}{8}\int  dx^0 dr d\alpha \left(
\frac{1}{y^3}\sqrt{\left(1+y'^2\right) \left(r^2+
y^2(\cos^2\xi\sin^2\xi\cos^2\theta_1
+\cos^2\xi\sin^2\theta_1)\dot{\phi}_1^2\right)} -
\frac{r}{y^3}\right) \ . \nonumber \\
\nonumber \\
\end{eqnarray}
In order to simplify the analysis we
note that the equation of motion  for
$\theta_1$ for non-zero $\xi$ has two
solution $\theta_1=0$ and
$\theta_1=\pi$ and we choose
$\theta_1=0$. Then the equation of
motion for $\xi$ has solutions for
$\xi=0,\frac{\pi}{2}$ or
$\xi=\frac{\pi}{4}$ and in order to
find non-trivial configuration we
choose $\xi=\frac{\pi}{4}$. Then the
action (\ref{SD21}) simplifies
considerably
\begin{eqnarray}\label{SD21s}
S_{D2}&=&  \tau_2  \frac{R^3}{8}\int
dx^0 dr d\alpha \left(
\frac{1}{y^3}\sqrt{\left(1+y'^2\right)
\left(r^2+
\frac{y^2}{4}\dot{\phi}_1^2\right)} -
\frac{r}{y^3}\right) \ . \nonumber \\
\end{eqnarray}
Then the equation of motion for $y$
takes the form
\begin{eqnarray}\label{eqy}
&-&\frac{3}{y^4}\sqrt{\left(1+y'^2\right)
\left(r^2+
\frac{y^2}{4}\dot{\phi}_1^2\right)}
-\frac{d}{dr} \left[\frac{y' \sqrt{r^2+
\frac{y^2}{4}\dot{\phi}_1^2}}
{y^3\sqrt{1+y'^2}}\right]+ \nonumber \\
&+&\frac{1}{4y^2}\frac{\sqrt{1+y'^2}\dot{\phi}_1^2}
{\sqrt{r^2+\frac{y^2}{4}\dot{\phi}_1^2}}+
\frac{3r}{y^4}=0 \
\nonumber \\
\end{eqnarray}
while  the equation of motion for
$\phi$ is equal to
\begin{equation}\label{eqphi1}
\frac{d}{d\alpha} \left[ \frac{
\sqrt{1+y'^2}\dot{\phi}_1} {y\sqrt{
r^2+
\frac{y^2}{4}\dot{\phi}_1^2}}\right]=0
\end{equation}
and it can be easily shown that the
equations (\ref{eqy}) and
(\ref{eqphi1}) can be solved with the
ansatz
\begin{equation}\label{ansD2}
y=\kappa r \ , \quad \phi_1=2\alpha \ .
\end{equation}
Similar half-BPS configuration was
previously studied in
\cite{Constable:2002xt} in the case of
$AdS_5\times S^5$ background. The
induced metric on the world-volume of
D2-brane is equal to
\begin{eqnarray}
ds^2=g_{MN}\partial_\alpha x^M
\partial_\beta X^N=
\frac{\tR^2}{\kappa^2}(1+\kappa^2)(d\xi^2)^2+
\frac{\tR^2}{\kappa^2 (\xi^1)^2}
[(d\xi^0)^2+(1+\kappa^2)(d\xi^1)^2]
\nonumber \\
\end{eqnarray}
that clearly shows that this D2-brane
configuration has a topology
$AdS_2\times S^1$.
Let us now evaluate the action on this ansatz
\begin{eqnarray}
S=\tau_{2}\frac{R^3}{8\kappa} 2\pi T \int_\epsilon^\infty
\frac{dr}{r^2}= \frac{R^3}{8 \kappa}2\pi T\frac{1}{\epsilon} \ ,
\nonumber \\
\end{eqnarray}
where $T$ is  regularized interval in
$x^0$ direction and where  we have
introduced a regulator $\epsilon << 1
$. It is however important to stress
that since we consider D2-brane that
has a finite extend we should take the
boundary terms into account. We will
discuss the boundary contributions in
more details bellow. Here we only
stress that,  following
\cite{Drukker:2005kx}, that  we should
add to the action the boundary
contribution in the form
\begin{eqnarray}
\delta S = 
-\left. y \frac{\delta \mL}{\delta y'}\right|_{y=\kappa\epsilon}
=-\tau_{2}\frac{R^3}{8}2\pi T \left.
y\frac{y'\sqrt{r^2+\frac{y^2}{4}\dot{\phi}^2}}
{y^3\sqrt{1+y'^2}}\right|_{y=\epsilon \kappa} =
-\tau_{2}\frac{R^3}{8}
2\pi T \frac{1}{\kappa\epsilon} \ .  \nonumber \\
\end{eqnarray}
 Consequently we obtain the
action after adding the boundary
contribution
\begin{equation}
S = S+\delta S=0 \ .
\end{equation}
We see that the action vanishes and hence the vacuum expectation
value of the dual line operator corresponding to the D2-brane
explained above, is equal to one. This is a strong indication that
the dual line operator is stable and preserves some fractions of
supersymmetry. It would be really interesting to study these
operators from the point of view of $3d$ QFT, following
\cite{Gukov:2008sn,Gukov:2006jk}.
\subsection{D2-brane with topology AdS$_2\times$ S$^1$}
In this section we present another
example of  D2-brane solution that has
a topology AdS$_2\times$ S$^1$. Once
again we start with the metric on
AdS$_4$ as
\begin{eqnarray}
ds^2_{AdS_4} = \frac{\tR^2}{ Y^2}[dY^2+(dX^\mu)^2], \>\>\> \mu =
0,1,2 .
\end{eqnarray}
This D2-brane can be described by the following parametrization:
\begin{eqnarray}
X^0 &=& \xi^0 \ , \quad  Y=\xi^1\equiv \sigma \ , \quad X^1\equiv
X(\sigma) \ , \quad \xi^2=\phi_1 \ , \cr & \cr \theta &=&
\theta_1=\theta_2=\psi= \phi_2 = {\rm const} \ , \quad
F_{01}=-F_{10}=F \ .
\end{eqnarray}
The DBI part of the D2-brane action is
written as \begin{equation}
 S_{DBI} =
\tau_2 e^{-\Phi_0}\int d^3 \zeta \sqrt{\det \bA} \ .
\end{equation}
Further, there is a WZ term in the
following form
\begin{eqnarray}
S_{WZ}= - i\tau_{2}\int F\wedge C^{(1)}
= -\frac{i\tau_{2}k}{2}\int d^3\zeta
(2\pi\alpha')
F_{01}\cos^2\xi\cos\theta_1 \ ,
\end{eqnarray}
where the non-zero one form R-R field
is
\begin{equation}
C^{(1)}_{\phi_1}=\frac{k}{2}\cos^2\xi
\cos\theta_1 \ .
\end{equation}
The fact that the WZ term in Euclidean
signature contains factor $i$ implies
that we have to consider imaginary
electric flux so that we can write
\begin{eqnarray}
\bA_{00}&=& \frac{\tR^2}{\sigma^2} \ ,
\quad
\bA_{11}=\frac{\tR^2}{\sigma^2}(1+X'^2)
\ , \quad
\bA_{01}=-\bA_{10}=i(2\pi\alpha')F \ , \nonumber \\
\bA_{22}&=&\tR^2(\cos^2\xi\sin^2\xi\cos^2\theta_1+
\cos^2\xi\sin^2\theta_1) \ ,
\end{eqnarray}
where $X' = \partial X / \partial \sigma $. Hence we obtain
\begin{equation}
S_{DBI} =\tau_2 \int {d^3 \zeta \tR^2\sqrt{\bA_{22}} \sqrt{
\frac{1}{\sigma^4}(1+X'^2)- \frac{(2\pi\alpha')^2}{\tR^4}F^2}} \ .
\end{equation}

  Let us now solve the equations
of motion coming from the above DBI
action including the WZ term. First of
all the variation of the action with
respect to $\theta_1$ implies
\begin{eqnarray}
& & \tau_{2}e^{-\Phi_0} \frac{\tR^3
\cos^4\xi \sin\theta_1\cos\theta_1}
{\sqrt{\sin^2\xi\cos^2\xi\cos^2\theta_1+\cos^2\xi
\sin^2\theta_1}}
\sqrt{\frac{1}{\sigma^4}(1+X'^2)-
\frac{(2\pi\alpha')^2}{\tR^4}F^2} -
\nonumber \\ &-&
\frac{\tau_{2}k}{2}(2\pi\alpha')F
\cos^2\xi\sin\theta_1=0 \ .
\end{eqnarray}
When $\cos \xi \neq 0$, the equation above is solved with
$\theta_1 = 0, \pi$ and we choose for our convenience
$\theta_1=0$. Further the equation of motion for $\xi$ implies
\begin{eqnarray}
\tau_{2}e^{-\Phi_0}
(\cos^2\xi-\sin^2\xi)
\tR^3\sqrt{\frac{1}{\sigma^4}(1+X'^2)-
\frac{(2\pi\alpha')^2F^2}{\tR^4}}-
\tau_{2}k(2\pi\alpha')F\cos\xi
\sin\xi=0 \nonumber \\
\end{eqnarray}
Now let us consider the equations of
motion for $A_0, A_1$ that again imply
the existence of a conserved electric
flux $\Pi$:
\begin{eqnarray}
e^{-\Phi_0}\frac{\tR^2}{\tR^4}
\sqrt{\bA_{22}}\frac{(2\pi\alpha')^2F}
{\sqrt{\frac{1}{\sigma^4}(1+X'^2)-\frac{(2\pi\alpha')^2F^2}
{\tR^4}}}-\frac{2\pi\alpha'k}{2}
\cos^2\xi=\Pi
\end{eqnarray}
Simplifying the above equations one gets:
\begin{eqnarray}
\sqrt{\frac{1}{\sigma^4}(1+X'^2)-
\frac{(2\pi\alpha')^2F^2}{\tR^4}}=
\frac{\sqrt{1+X'^2}}
{\sigma^2\sqrt{1+\frac{e^{2\Phi_0}}{\tR^2\sin^2\xi}
(\frac{\Pi}{2\pi\alpha'\cos\xi}+\frac{k}{2}\cos\xi)^2}}
\nonumber \\
\end{eqnarray}
Inserting the above result into the
equation of motion for $\xi$ we obtain
\begin{eqnarray}
\frac{\Pi}{2\pi\alpha'}=-\frac{k}{4} \
.
%
\end{eqnarray}
Note that this equation is obeyed for
any $\xi$.
 Notice also that in deriving the above, we
have used the fact that
\begin{eqnarray}
\frac{e^{2\Phi_0}}{\tR^2}=\frac{4}{k^2}  \ .
\nonumber \end{eqnarray}
Finally we consider equation of motion for $X$ that takes the
following simple form
\begin{eqnarray}
\partial_\sigma[\frac{X'}{\sigma^4
\sqrt{(\dots)}}]=0 \ .
\end{eqnarray}
It is clear that this equation has
natural solution $X'=0\Rightarrow X=
{\rm const.}$. Then the metric induced
on the world-volume of D2-brane takes
the form
\begin{equation}
ds_{in}^2=g_{mn}\partial_\alpha X^m
\partial_\beta X^n d\xi^\alpha
d\xi^\beta=\frac{\tR^2}{\sigma^2}
((d\xi^0)^2+d\sigma^2)+\tR^2\sin^2\xi
\cos^2\xi (d\xi_2)^2 \
\end{equation}
that clearly has a form $AdS_2\times
S^1$.

 Let us again compute the current
$J^{01}$ which takes the form
\begin{eqnarray}
J^{01}= -\tau_{2}\frac{1}{2\pi\alpha'}
\left(\Pi+\frac{2\pi\alpha'
k}{2}\cos^2\xi \right)+
\frac{\tau_{2}}{2} k\cos^2\xi =
-\tau_{2} \frac{\Pi}{(2\pi\alpha')}  =
\frac{\tau_2 k}{4} \ .
 \nonumber \\
\end{eqnarray} Note that it is
again proportional to the number of fundamental strings $k$.

Let us now evaluate the action on the solution given above
\begin{eqnarray}
S &=& \tau_{2}e^{-\Phi_0}\tR^3
\int_0^{2\pi} d\phi
\int_{-T/2}^{T/2}d\xi^0
\int_\epsilon^\infty d\sigma
\sin\xi\cos\xi
\sqrt{\frac{1}{\sigma^4}(1+X'^2)-
\frac{(2\pi\alpha'F)^2}{\tR^4}}+\nonumber \\
&+& \frac{\tau_{2}}{2}k \int_0^{2\pi}
\int_{-T/2}^{T/2}d t
\int_\epsilon^\infty d\sigma
(2\pi\alpha')F\cos^2\xi
\nonumber \\
&=& \frac{2\pi T\tau_{2}e^{\Phi_0}\tR}
{\sin\xi\cos\xi}
\left(\frac{k^2}{4}\sin^2\xi+k\frac{\Pi}{4\pi\alpha'}+
\frac{k^2}{4}\cos^2\xi\right)\cos^2\xi
\times \nonumber \\
&\times& \int_\epsilon^\infty d\sigma
\frac{ \sqrt{1+X'^2}}{\sigma^2
\sqrt{1+\frac{e^{2\Phi_0}}{\tR^2\sin^2\xi}
(\frac{\Pi}{2\pi\alpha'\cos\xi}+\frac{k}{2}\cos\xi)^2}}
\nonumber \\
&=& \frac{2\pi T\tau_{2}e^{\Phi_0}\tR
k^2}{4}\cos^2\xi\frac{1}{\epsilon} \ .
\nonumber \\
\end{eqnarray}
Where deriving above we have used the following identities
\begin{eqnarray}
\frac{e^{2\Phi_0}}{\tR^2}=\frac{4}{k^2}
\ , \quad
1+\frac{e^{2\Phi_0}}{\tR^2\sin^2\xi
\cos^2\xi}
\left(\frac{\Pi}{2\pi\alpha'}+
\frac{k}{2}\cos^2\xi\right)^2=\frac{1}{4\sin^2\xi\cos^2\xi}
\end{eqnarray}
Notice that this action derived above
is proportional to the world volume
electric flux $\Pi$.
\subsection{Boundary terms}
The D2-brane solution that we found corresponds to D2-brane that
extends all the way to the boundary of AdS$_4$ and ends there
along a one-dimensional curve. Since this D2-brane action has
finite extent we have to discuss the possibility of adding
boundary terms to the action. These boundary terms should not
change the equations of motion and hence the  solution will still
be the same, but the value of the action when evaluated at this
solution will in general depend on the boundary terms.

As it is well known that when we
calculate the Wilson loop using string
surfaces the bulk action is divergent
but this divergence can be canceled by
boundary term as we will review in the
Appendix. Explicitly, the string that
corresponds to Wilson loop has to
satisfy three Dirichlet boundary
conditions  on the three directions
parallel to the boundary of AdS$_4$ ,
and the seven Neumann ones that combine
the radial coordinate of AdS$_4$ and
coordinates along $CP_3$ and hence it
is the appropriate action assuming that
we have Dirichlet boundary conditions.
So we have to add appropriate boundary
terms that change the boundary
conditions.

In case of the fundamental string we
have the coordinate $Y$ with Neumann
boundary condition (before imposing the
static gauge). Then it is natural to
define $p_y$ as the momentum conjugate
to it
\begin{equation}
p_Y = \frac{\delta S}{\delta \partial_n
Y} \ ,
\end{equation}
where $n$ is normal derivative to the boundary. The new action
including the term that changes the boundary conditions is
\begin{equation}
\tilde{S} = S - Y_0\int d\tau p_Y
\end{equation} where the integral is
over the boundary at a cutoff $Y =
Y_0$. In fact, the variation of the
original action is
\begin{eqnarray}
\delta S= \int d^2\sigma \left[\frac{\delta \mL}{\delta
Y}-\partial_\alpha \frac{\delta \mL}{\delta\partial_\alpha
Y}\right]\delta Y + \int d\tau \frac{\delta \mL}{\delta
\partial_\sigma Y}\delta Y=
\int d\tau \frac{\delta \mL}{\delta
\partial_\sigma Y}\delta Y \ ,
\nonumber \\
\end{eqnarray}
where we used the fact that the field
$Y$ obeys the equation of motion in the
bulk. The boundary term above shows
(since it is proportional to $\delta
Y$) that on-shell action is functional
of $Y$. Then including the boundary
term we obtain
\begin{equation}
\delta \tilde{S}= \int d\tau [ p_Y
\delta Y-\delta p_Y Y-p_Y\delta Y]=
-\int d\tau \delta p_Y Y_0
\end{equation}
and hence the new action is functional
of $p_Y$ as it should be.

Taking a lesson from the review of the boundary term for the
fundamental string mentioned above we now return to D2-brane
action. In case of D2-brane action there is another subtlety. The
DBI action is a functional of the gauge field, but the Wilson loop
observable should depend on the variable $\Pi$. Then in order to
find correct form of the action we have to add to it following
boundary term
\begin{eqnarray}
\delta S &=& -\int dt d\phi \left[\frac{\delta \mL}{\delta
\partial_\sigma Y}Y
+
\frac{\delta \mL}{\delta
\partial_\sigma
A_0}A_0\right] =\nonumber \\
&=& -2\pi T\frac{\delta \mL}{\delta
\partial_\sigma Y}Y
 +2\pi T \tau_{2}
\int_\epsilon^\infty d\sigma \Pi
F_{0\sigma}=  \nonumber \\
&=& - \tau_{2}e^{\Phi_0} 2\pi T
\tR\frac{k^2}{4}
\cos^2\xi\frac{1}{\epsilon} \ ,
\end{eqnarray}
where $\frac{\delta \mL}{\delta
\partial_\sigma A_0}=-\Pi={\rm const}.$
Now if we collect all these terms together we obtain
\begin{eqnarray}
S + \delta S &=&0
\end{eqnarray}
We  find the straight Wilson line has
vanishing action after adding the
boundary term which is the same result
as in case of $AdS_5/CFT_4$
correspondence.

\subsection{Circular Wilson line}
In order to find this solution we consider following form of
AdS$_4$ metric
\begin{equation}
ds^2_{AdS_4}=\frac{1}{y^2}(dy^2+dr^2+r^2d\phi^2+d(x^0)^2)
\end{equation}
and consider the Wilson line that is siting at $x^0=0$.
Our goal is to find Wilson line that for $y=0$
takes the form of circle with $r^2=R^2$. In order
to find such a configuration we consider an ansatz
\begin{eqnarray}
\sigma^0&=&\phi \ , \quad
y=\sigma^1\equiv \sigma \ ,
\quad r=r(\sigma) \ , \nonumber \\
F_{01}&=& -F_{10}=iF \ , \quad
\phi^2=\xi^2 \ , \quad
\theta_1=\theta_2=\psi=\phi_1= {\rm
const} \ .
\end{eqnarray}
Then we easily obtain
\begin{eqnarray}
\bA_{00}&=&\tR^2\frac{r^2}{y^2} \ , \quad
\bA_{11}=\tR^2\frac{1}{y^2}(1+r'^2) \ , \nonumber \\
\bA_{01}&=&i(2\pi\alpha')F \ , \quad
\bA_{10}=-i(2\pi\alpha')F \ , \nonumber \\
\bA_{22}&=&\tR^2(\cos^2\xi\sin^2\xi\cos^2\theta_1+
\cos^2\xi\sin^2\theta_1) \ , \nonumber \\
\end{eqnarray}
where $r'=\partial_\sigma r$. Then we
obtain
\begin{equation}
\det\bA=
\tR^4\bA_{22}(\frac{r^2}{y^4}(1+r'^2)-\frac{(2\pi\alpha')^2F^2}{\tR^4})
\end{equation}
so that DBI part of the action takes the form
\begin{equation}
S=\tau_{2}e^{-\Phi_0}\tR^2 \int
d^2\sigma dy \sqrt{\bA_{22}}\sqrt{
\frac{r^2}{y^4}
(1+r'^2)-\frac{(2\pi\alpha')^2F^2}{\tR^4}}
\ .
\end{equation}
Further, there is  a coupling to
$C^{(1)}$ field in the form
\begin{equation}
S_{WZ}=-i\tau_{2}\int (2\pi\alpha')F\wedge C=
-i\frac{\tau_2(2\pi\alpha')k}{2} \int d^3\zeta
F_{01}\cos^2\xi\cos^2\theta_1 \ .
\end{equation}
As in the case of straight Wilson line we start with the equation
of motion for $\theta_1$
and we found that it is solved for
$\theta_1 = 0$, or $\theta_1 = \pi$ and
we choose for our convenience
$\theta_1=0$. Further the equation of
motion for $\xi$ implies
\begin{eqnarray}
\tau_{2}e^{-\Phi_0}
(\cos^2\xi-\sin^2\xi)
\tR^3\sqrt{\frac{r^2}{y^4}(1+r'^2)-
\frac{(2\pi\alpha')^2F^2}{\tR^4}}-
\tau_{2}k(2\pi\alpha')F\cos\xi
\sin\xi=0 \ .  \nonumber \\
\end{eqnarray}
Further, the equations of motion for
$A_0, A_1$ imply  a conserved electric
flux $\Pi$
\begin{equation}
e^{-\Phi_0}\frac{\tR^2}{\tR^4}
\sqrt{\bA_{22}}\frac{(2\pi\alpha')^2F}
{\sqrt{\frac{r^2}{y^4}(1+r'^2)-\frac{(2\pi\alpha')^2F^2}
{\tR^4}}}-\frac{2\pi\alpha'k}{2}
\cos^2\xi=\Pi
\end{equation}
that allows us to find
\begin{eqnarray}
& &\frac{2\pi\alpha'F}{\tR^2}
=\frac{e^{\Phi_0}}{\tR\sin\xi}
(\frac{\Pi}{2\pi\alpha'\cos\xi}+\frac{k}{2}\cos\xi)
\frac{r\sqrt{1+r'^2}}{
y^2\sqrt{1+\frac{
e^{2\Phi_0}}{\tR^2\sin^2\xi}
(\frac{\Pi}{2\pi\alpha'\cos\xi}+\frac{k}{2}\cos\xi)^2}}
\ ,
\nonumber \\
& &\sqrt{\frac{r^2}{y^4}(1+r'^2)-
\frac{(2\pi\alpha')^2F^2}{\tR^4}}=
\frac{r\sqrt{1+r'^2}}
{y^2\sqrt{1+\frac{e^{2\Phi_0}}{\tR^2\sin^2\xi}
(\frac{\Pi}{2\pi\alpha'\cos\xi}+\frac{k}{2}\cos\xi)^2}}
\ .
\nonumber \\
\end{eqnarray}
Inserting the above results into the
equation of motion for $\xi$ we again
obtain
\begin{eqnarray}
\frac{\Pi}{2\pi\alpha'}=-\frac{k}{4} \
.
\end{eqnarray}
Note that in deriving the above, we have used the fact that
\[\frac{e^{2\Phi_0}}{\tR^2}=\frac{4}{k^2} \ . \]
Finally we determine the equation of
motion for $r$
\begin{eqnarray}
\frac{1}{y^2}\sqrt{1+r'^2}
-\frac{d}{d\sigma} \left[\frac{rr'}{y^2
\sqrt{1+r'^2}}\right]=0 
\end{eqnarray}
Now we will argue that the ansatz $r^2=R^2-y^2$ solves the
equation above. Indeed, using the fact that $r'=-\frac{y}{r}$ we
one can check that the above equation is identically zero.
Let us now evaluate the action for the D2-brane configuration on
this solution
\begin{eqnarray}
S=& &\tau_{2}e^{-\Phi_0} \int d\phi
d\phi^2 dy\tR^3 \sin \xi \cos\xi
\sqrt{\frac{r^2}{y^4}(1+r'^2)+
\frac{(2\pi\alpha')^2F^2}{\tR^4}}
+\nonumber \\
&+&\frac{\tau_{2} k\tR^2}{2} \int d\phi
d\phi^2 dy \frac{(2\pi\alpha')F}{\tR^2}
\cos^2\xi=\nonumber \\
 &=& 4\pi^2\tau_{D2}e^{\Phi_0}\tR
\frac{k^2}{4}\cos^2\xi
R[-\frac{1}{R}+\frac{1}{\epsilon}] \ .
\nonumber \\
\end{eqnarray}
Further, we have the first boundary
contribution at $\sigma=\epsilon$.
We again proceed as in previous section for the boundary
contributions. Namely, we evaluate the contribution to the action
for general $y$, then insert the ansatz $y=\sigma$ and finally
evaluated the action at $y=\epsilon$
\begin{eqnarray}
{\delta S}_Y = -\int d\phi^2
d\phi\left[ \frac{\delta \mL}{\delta
\partial_\sigma y}y\right]
= -4\pi^2\tau_{2}e^{\Phi_0}\tR
\frac{k^2}{8} \frac{R^2-\epsilon^2}
{\epsilon R} \ . \nonumber \\
\end{eqnarray}
In the same way the boundary
contribution from the gauge fields
takes the form
\begin{eqnarray}
{\delta S}_{A} &=& \tau_{2}\int d\phi
d\phi^2 \int_\epsilon^R d\sigma \Pi
F_{\tau\sigma} =
\nonumber \\
&=& 4\pi^2\tau_{2}\tR e^{\Phi_0}
\frac{k^2}{16}(1-2\cos^2\xi)
\frac{2R}{1} \int_\epsilon^R
\frac{d\sigma}{\sigma^2}
\nonumber \\
&=& 4\pi^2\tau_{2}\tR e^{\Phi_0}
\frac{k^2R}{8}(1-2\cos^2\xi)
\left[-\frac{1}{R}
+\frac{1}{\epsilon}\right] \ .  \nonumber \\
\end{eqnarray}
Collecting all these terms together we obtain that the divergent
terms cancel as in the case of straight Wilson line. On the other
hand we find finite contribution to the action in the form
\begin{eqnarray}
S + {\delta S}_Y + {\delta S}_{A_0} =
-4\pi^2\tau_{2}e^{\Phi_0}\tR\frac{k^2}{8}=
\frac{k}{4}S_{FS} \ ,
\end{eqnarray}
where we used the convention that
$\tau_{2}=\frac{1}{4\pi^2}$ and  where
$S_{FS}$ is the fundamental string
action evaluated on circular Wilson
line. Borrowing the interpretation of
Wilson lines in $AdS_5/CFT$
correspondence using D3-branes we can
argue that our solution describes
Wilson line in symmetric $\frac{k}{4}$
representation where $k$ is level of CS
action. It would be certainly very
interesting to study the problem
whether there exists D-brane
description of Wilson loops in
arbitrary representations. We hope to
return to these problems in future.

\section{Conclusions} \label{fifth}
We have studied in this paper various
D-brane configurations corresponding to
Wilson loops and lines in the boundary
of AdS$_4$. We have taken D2-brane as
an example and have shown various
configurations of this in the Euclidean
AdS$_4 \times$ CP$^3$ correspond to
straight and circular Wilson line
solutions.  We  analyzed the D2-brane
which has a topology of AdS$_2 \times$
S$^1$ and corresponds to straight
Wilson line gives vanishing action
after adding the boundary term. We have
also studied the circular Wilson line
solution and show that the action gives
a non-zero contribution after adding
the boundary term. One can study the
D4-brane which correspond to both
straight and circular Wilson line in
the boundary AdS$_4$, however we
mention that the action will take a
similar form like that of D2-brane and
hence the analysis is very similar to
the one performed in the present paper.
\\
\\
Note added: While we were finalizing to
submit our paper, we came across
\cite{Drukker:2008zx} and
\cite{Chen:2008bp} which have some
overlap with our present work.
\\
\vskip .2in \noindent {\bf Acknowledgements:} We would like to
thank an anonymous referee for finding out mistakes in the first
version of the manuscript. KLP would like to thank the hospitality
at Institute of Physics, Bhubaneswar, India where apart of this
work was done. This work was partially supported by the Science
Research Center Program of the Korea Science and Engineering
Foundation through the Center for Quantum Spacetime (CQUeST) of
Sogang University with grant number R11 - 2005 - 021. The work of
JK  was supported by the Czech Ministry of Education under
Contract No. MSM 0021622409.

\end{document}